\documentclass[preprint]{vgtc}               

\ifpdf
  \pdfoutput=1\relax                   
  \usepackage{graphicx}                
  \DeclareGraphicsExtensions{.pdf,.png,.jpg,.jpeg} 
\else
  \usepackage{graphicx}                
  \DeclareGraphicsExtensions{.eps}     
\fi%

\graphicspath{{figures/}{pictures/}{images/}{./}} 

\usepackage{microtype}                 
\PassOptionsToPackage{warn}{textcomp}  
\usepackage{textcomp}                  
\usepackage{mathptmx}                  
\usepackage{times}                     
\usepackage{cite}                      
\usepackage{tabu}                      
\usepackage{booktabs}                  

\usepackage{multirow}
\usepackage{afterpage}
\usepackage{adjustbox}
\usepackage{amsmath}
\DeclareMathOperator{\boldx}{\mathbf{x}}

\newcommand\nObj{100,000 }
\usepackage{balance}
\usepackage{appendix}
\usepackage[position=bottom]{subfig}

\usepackage{array}
\usepackage{graphicx}
\usepackage{amsfonts}
\usepackage{enumitem}
\usepackage{soul}
\usepackage{hhline}
\usepackage{multirow, makecell}
\usepackage{float}
\usepackage{booktabs}

\usepackage{amsthm}
\usepackage{color}
\usepackage{transparent}
\usepackage{url}
\usepackage{footmisc}
\usepackage{setspace}
\usepackage{mathtools}
\usepackage{xcolor}
\usepackage{tabularx}
\definecolor{lg}{HTML}{9bfaa8}
\DeclareRobustCommand{\hllg}[1]{#1}

\usepackage{amssymb}
\usepackage{pifont}

\usepackage{svg}
\usepackage{adjustbox}

\newif\ifcomment
\def\comment#1{%
    \ifcomment\relax\else #1\fi}
\commenttrue

\newcommand\blfootnote[1]{%
  \begingroup
  \renewcommand\thefootnote{}\footnote{#1}%
  \addtocounter{footnote}{-1}%
  \endgroup
}

\onlineid{2639}

\vgtccategory{Research}

\vgtcinsertpkg

\preprinttext{}

\title{Learning Acoustic Scattering Fields for Dynamic Interactive Sound Propagation}

\author{Zhenyu Tang\thanks{Equal contribution}, Hsien-Yu Meng\footnotemark[1], and Dinesh Manocha}
\authorfooter{
\item
 Zhenyu Tang, Hsien-Yu Meng, and Dinesh Manocha are with the University of Maryland. E-mail: \{zhy, mengxy19, dm\}@umd.edu.
}

\teaser{
  \centering
  \includegraphics[width=\linewidth]{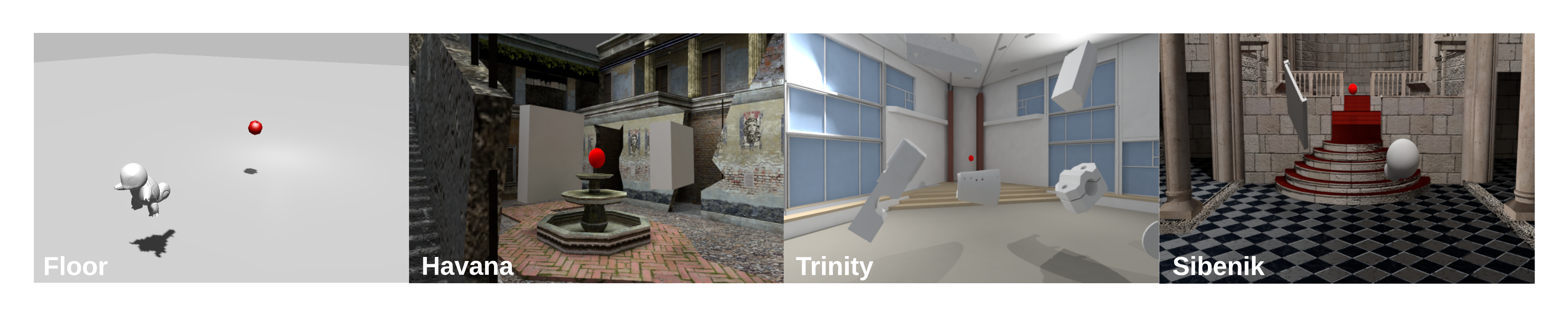}
  \caption{We show the dynamic scenes with various moving objects that are used to evaluate our hybrid sound propagation algorithm. We compute the acoustic scattered fields of each object using a neural network and couple them with interactive ray tracing to generate diffraction and occlusion effects. Our approach can generate plausible acoustic effects in dynamic scenes in a few milliseconds and we demonstrate its benefits for sound rendering in virtual environments.}
	\label{fig:teaser}
}

\abstract{We present a novel hybrid sound propagation algorithm for interactive applications. Our approach is designed for dynamic scenes and uses a neural network-based learned scattered field representation along with ray tracing to generate specular, diffuse, diffraction, and occlusion effects efficiently. We use geometric deep learning to approximate the acoustic scattering field using spherical harmonics. We use a large 3D dataset for training, and compare its accuracy with the ground truth generated using an accurate wave-based solver. The additional overhead of computing the learned scattered field at runtime is small and we demonstrate its interactive performance by generating plausible sound effects in dynamic scenes with diffraction and occlusion effects. We demonstrate the perceptual benefits of our approach based on an audio-visual user study. %
} 

\nocopyrightspace

\begin{document}



\maketitle

\section{Introduction}
Interactive sound propagation and rendering are increasingly used to generate plausible sounds that can improve a user's sense of presence and immersion in virtual environments~\cite{larsson2002}. Recent advances in geometric and wave-based simulation methods have lead to integration of these methods into current games and virtual reality (VR) applications to generate plausible acoustic effects, including Project Acoustics~\cite{MicrosoftAcoustics}, Oculus Spatializer~\cite{OculusS}, and Steam Audio~\cite{SteamA}. The underlying propagation algorithms are based on using reverberation filters~\cite{valimaki2012fifty}, ray tracing~\cite{schissler2014high,schissler2018interactive}, or precomputed wave-based acoustics~\cite{nikunj2014}. 

A key challenge in interactive sound rendering is handling  dynamic scenes that are frequently used in games and VR applications. Not only can the objects undergo large motion or deformation, but their topologies  may also change. In addition to specular and diffuse effects, it is also important to simulate complex diffracted scattering, occlusions, and inter-reflections that are perceptible~\cite{james2006precomputed,pulkki2019machine,nikunj2014}. Prior geometric methods are accurate in terms of simulating high-frequency effects and can be augmented with approximate edge diffraction methods that may work well in certain cases~\cite{tsingos2001,schissler2014high}, though their behavior can be erratic~\cite{rungta2016psychoacoustic}. On the other hand, wave-based precomputation methods can accurately simulate these effects, but are limited to static scenes~\cite{nikunj2014,raghuvanshi2018parametric}. Some hybrid methods are limited to interactive dynamic scenes with well-separated rigid objects~\cite{rungta2018diffraction}. Our goal is to design similar hybrid methods that can overcome these restrictions and can  generate diffraction and occlusion effects  that translate into good perceptual differentiation~\cite{rungta2016psychoacoustic}.


Many recent works use machine learning techniques for audio processing, including recovering acoustic parameters of real-world scenes from recordings~\cite{eaton2016estimation,genovese2019blind,tsokaktsidis2019artificial}. Furthermore, learning methods have been used to approximate diffraction scattering and occlusion effects from rectangular plate objects~\cite{pulkki2019machine} and frequency-dependent loudness fields for 2D convex shapes~\cite{fan2019fast}. These results are promising and have motivated us to develop good learning based methods for more general 3D objects.

\noindent{\bf Main Results:} We present a novel approach to approximate the acoustic scattering field of an object in 3D using neural networks for interactive sound propagation in dynamic scenes. Our approach makes no assumption about the motion or topology of the objects. We exploit properties of the acoustic scattering field of objects for lower frequencies and use neural networks to learn this field from geometric representations of the objects.
\blfootnote{Our code and data will be released after publication.}
Given an object in 3D, we use the neural network to estimate the scattered field at runtime, which is used to compute the propagation paths when sound waves interact with objects in the scene. The radial part of the acoustic scattering field is estimated using geometric ray tracing, along with specular and diffuse reflections. Some of the novel components of our work include:
\begin{itemize}
    \item {\bf Learning acoustic scattering fields:} We use techniques based on geometric deep learning
    to approximate the angular component of acoustic wave propagation in the wave-field. 
    Our neural network takes the point cloud as the input and outputs the spherical harmonic coefficients that represent the acoustic scattering field. We compare the accuracy of our learning method with an exact BEM solver, and the error on new, unseen objects (as compared to training data). Our empirical results are promising and we observe average normalized reproduction error\cite{lilis2010sound,betlehem2005theory} of  $8.8\%$ in the pressure fields.  
    \vspace*{-0.07in}
    \item {\bf Interactive wave-geometric sound propagation:} We present a hybrid propagation algorithm that uses a neural network-based scattering field representation along with ray tracing to efficiently generate specular, diffuse, diffraction, and occlusion effects at interactive rates.
    \vspace*{-0.07in}
    \item {\bf Plausible sound rendering for dynamic scenes:} We present the first interactive approach for plausible sound rendering in dynamic scenes with diffraction modeling and occlusion effects. As the objects deform or change topology, we compute a new spherical harmonic representation using the neural network. Compared with prior interactive  methods, we can handle unseen objects at real-time, without using precomputed transfer functions for each object.
    \vspace*{-0.07in}
    \item {\bf Perceptual evaluation:} We perform a user study to validate the perceptual benefits of our method. Our propagation algorithm generates more smooth and realistic sound and has increased perceptual differentiation over prior methods used for dynamic scenes~\cite{schissler2017interactive,rungta2018diffraction}.
    \end{itemize}
 We demonstrate the performance in dynamic scenes with multiple moving objects and changing topologies. The additional runtime overhead of estimating the scattering field from neural networks is less than $1$ms per object on a NVIDIA GeForce RTX 2080 Ti GPU. The overall running time of sound propagation is governed by the underlying ray tracing system and takes few milliseconds per frame on multi-core desktop PC. We also evaluate the accuracy of acoustic scattering fields, as shown in Figure~\ref{fig:heatmap}.
    
\section{Related Work}

\subsection{Sound Propagation}
Wave-based techniques to model sound propagation  solve the acoustic wave equation directly using numerical solvers such as the finite-element method~\cite{thompson2006review}, the boundary-element method~\cite{wrobel2003boundary}, the finite-difference time domain~\cite{botteldooren1995finite}, adaptive rectangular decomposition~\cite{raghuvanshi2009efficient}, etc. Their complexity increases linearly with the size of the environment (surface area or volume) and as a third or fourth power of frequencies. As a result, they are limited to lower frequencies~\cite{Raghuvanshi:2010:PWS,mehra2013wave,yeh2013wave}.

Geometric techniques model the acoustic effects based on ray theory and typically work well for high-frequency sounds to model specular and diffuse reflections~\cite{krokstad1968,lauterbach2007,savioja2015overview,funkhouser1998}.  These techniques can be enhanced to simulate low-frequency diffraction effects. This includes  the accurate time-domain  Biot-Tolstoy-Medwin (BTM) model, which can be expensive and is limited to offline computations~\cite{svensson1999}. For interactive applications, commonly used techniques are based on the uniform theory of diffraction (UTD), which is a less accurate frequency-domain model  that can generate plausible results in some cases~\cite{tsingos2001,Taylor12,schissler2014high}.  Moreover, the complexity of edge-based diffraction algorithms can increase exponentially with the maximum diffraction order.

\subsection{Interactive Sound Rendering in Dynamic Scenes}
At a broad level, techniques for dynamic scenes can be classified into reverberation filters, geometric and wave-based methods, and hybrid combinations. The simplest and lowest-cost algorithms are based on artificial reverberators~\cite{valimaki2012fifty}, which  simulate the decay of sound in rooms. These filters are designed based on different parameters and are  either specified by an artist or computed using scene characteristics~\cite{tsingos2009precomputing}. They can handle dynamic scenes but assume that the reverberant sound field is diffuse, making them unable to generate directional reverberation or time-varying effects.

Many interactive techniques based on geometric acoustics and ray tracing have been proposed for dynamic scenes~\cite{vorlander1989,Taylor12,schissler2017interactive}.  They use spatial data structures along with multiple cores on commodity processors and caching techniques to achieve higher performance. Furthermore, 
hybrid combinations of ray tracing and reverberation filters~\cite{schissler2018interactive} have been proposed for low-power, mobile devices. In practice, these methods can handle scenes with a large number of moving objects, along with sources and the listener, but can't model diffraction or occlusion effects well.  

Many precomputation-based wave acoustics techniques tend to compute a global representation of the acoustic pressure field. They are limited to static scenes, but can  handle real-time movement of both sources and the listener~\cite{Raghuvanshi:2010:PWS,mehra2015wave}. These representations are computed based on uniform or adaptive sampling techniques~\cite{chaitanya2019adaptive}.
Overall, the acoustic wave field is a complex high-dimensional function and many efficient techniques have been designed to encode this field~\cite{nikunj2014,raghuvanshi2018parametric} within $100$MB and with a small runtime overhead. A hybrid combination of BEM and ray tracing has been presented for dynamic scenes with well-separated rigid objects~\cite{rungta2018diffraction}. 
A recent \emph{Planeverb} system~\cite{rosen} is able to perform 2D  wave simulation at interactive rates and calculate perceptual acoustic parameters that can be used for sound rendering. 

\subsection{Machine Learning and Acoustic Processing}
Machine learning techniques are increasingly used for acoustic processing applications. These include isolating the source locations in multipath environments~\cite{ferguson2018sound} and recovering the room acoustic parameters corresponding to reverberation time, direct-to-reverberant ratio, room volume, equalization, etc. from recorded signals~\cite{eaton2016estimation,genovese2019blind,tsokaktsidis2019artificial,tang2019scene}. These parameters are used for speech processing or audio rendering in real-world scenes. Neural networks have also been used to replace the expensive convolution operations for fast auralization~\cite{tenenbaum2019room}, to render the acoustic effects of scattering from rectangular plate objects  for VR applications~\cite{pulkki2019machine}, or to learn the mapping from convex shapes to the frequency dependent loudness field~\cite{fan2019fast}. The last method formulates the scattering function computation as a high-dimension image-to-image regression and is mainly limited to convex objects that are isomorphic to spheres.
 

\section{Background and Overview}
\begin{figure*}[htbp]
  \includegraphics[width=0.95\linewidth]{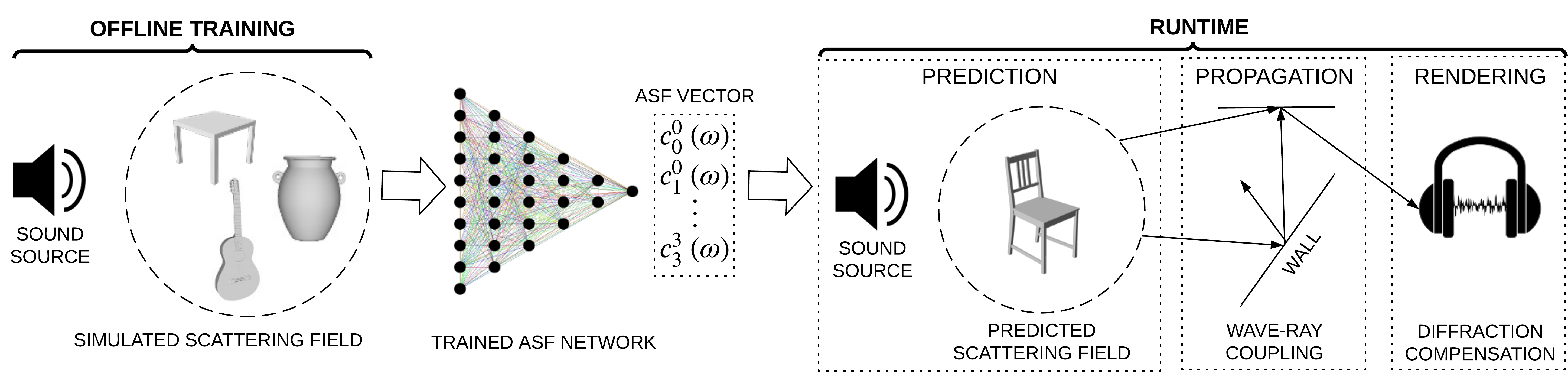}
  \caption{{\bf Overview:} Our algorithm consists of the training stage and the runtime stage. The training stage uses a large dataset of 3D objects and their associated acoustic pressure fields computed using an accurate BEM solver to train the network. The runtime stage uses the trained neural network to predict the sound pressure field from a point cloud approximation of different objects at interactive rates. 
  }
  \label{fig:overview}
 \vspace{-1em}
\end{figure*}

\subsection{Global and Localized Sound Fields}
Sound fields typically refer to the sound energy/pressure distribution over a bounded space as generated by one or more sound sources.  The global sound field in an acoustic environment depends on each sound source location, the propagating medium, and any reflections from boundary surfaces and objects.  This requires solving the wave equation in the free-field condition and evaluating inter-boundary interactions of sound energy using a global numeric solver (details in Appendix~\ref{sec:append_helmholtz}). In this case, the position of all scene objects/boundaries and sound sources needs to be specified beforehand, and any change in these conditions changes the sound field. The exact computation of the global pressure field is very expensive and can takes tens of hours on a cluster~\cite{mehra2013wave,Raghuvanshi:2010:PWS,nikunj2014}.

Our goal is to generate plausible sounds in virtual environments with dynamic objects. Therefore, it is important to model the acoustic scattering field (ASF) of each object. The ASFs of different objects are used to represent the localized pressure field, which is needed for diffraction and inter-reflection effects~\cite{james2006precomputed,mehra2013wave}. At the same time, the sound field in the free space (e.g., the far-field) between two distant objects is approximated using ray tracing, and we do not compute that pressure field accurately using a wave-solver. In practice, computing the sound field in a localized space for each object in the scene is much simpler and easier to represent than using a global solver~\cite{mehra2013wave,rungta2018diffraction}.

\subsection{Overview}

We present a learning method to approximate the ASFs of static or dynamic 3D objects of moderate sizes. 
In terms of correlation between the object shape and its scattering field, the volume of the scatterer closely relates to its low-order shape characteristics that can be represented by coarse triangle faces, which dominate the low-frequency scattering behaviors; while at high frequencies, this relationship shifts to high-order shape characteristics (i.e., geometrical details). Given the powerfulness of deep learning inference, we hypothesize the scattering sound distribution can be directly learned from the scatterer geometry, without solving the complicated wave equations. The inference speed on a modern GPU far exceeds conventional wave solvers, making deep neural networks suitable for interactive sound rendering applications. Therefore, we propose using appropriate 3D representation of objects to feed a neural network that can learn its corresponding scattered acoustic pressure field. We build and evaluate our method mainly on low frequency sounds and leverage state-of-the-art geometric ray-tracing techniques to handle high frequency sounds.

For each object, we consider a spherical grid of incoming directions and  model the plane-waves from each direction of this grid. For each plane wave, our goal is to compute the scattered field for the object on an offset surface of the object. Our geometric deep learning method is used to compute the angular portion of the scattered field (Equation~\ref{eq:solution}). If two objects move and are in a touching configuration, our learning algorithm treats them as a one large object and estimates its scattered field. Similarly, we can recompute the scattered field for a deforming object. 
An overview of our approach is illustrated in Figure~\ref{fig:overview}. 

\section{Learning-based Sound Scattering}

\subsection{Wave Propagation Modeling}
\label{sec:wave_modeling}

Our approach is designed for synthetic scenes and we assume a geometric representation (e.g., triangle mesh) is given to us. 
So the acoustic scattering field $p(\boldx, \omega)$ around the object can be solved numerically (derivation in Appendix~\ref{sec:append_helmholtz} and~\ref{sec:append_scattering}). 
In this work, we propose modeling the angular part of the scattering field using our learning based pressure field inference. 
The radial part is approximated using geometric sound propagation techniques.

\subsubsection{Radial Decoupling}
\begin{figure}[htbp]
\centering
  \includegraphics[width=0.8\linewidth]{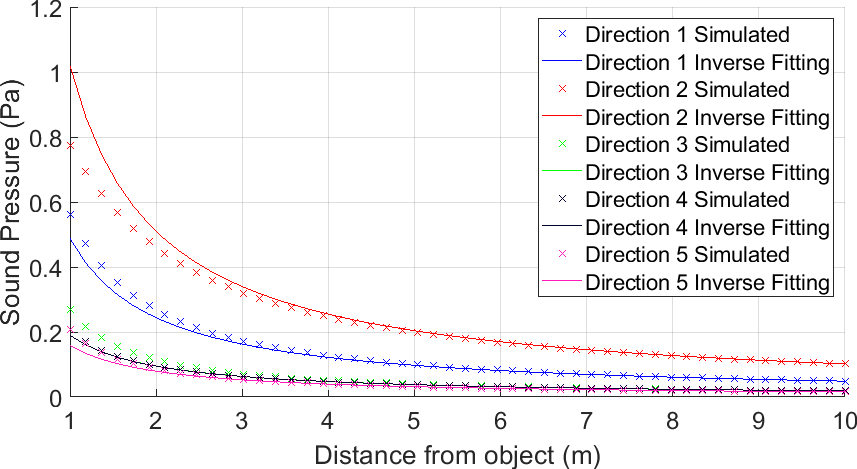}
  \caption{{\bf Simulated sound pressure fall-off and inverse-distance law fitted curves:} We calculate the sound pressure around a sound scatterer in our dataset using the BEM solver as reference. We examine the sound pressure from $1m$ to $10m$ scattered along 5 directions ($0^\circ, 72^\circ, 144^\circ, 216^\circ$, and $288^\circ$). We regard the sound pressure value at $10m$ to correspond to far-field condition, and inversely fit the pressure values for distance within $10m$ according to Equation~\ref{eq:inversedist}. We use$r_{ref}=5m$ is used for generating our ASFs, although other values can be used as well. }
  \label{fig:inverse}
  \vspace{-0.1in}
\end{figure}

Our goal is to determine the scattering field over the exterior space $E$ using a wave-solver. This field needs to be compactly encoded for efficient training. As shown in Equation~(\ref{eq:solution}), acoustic wave propagation in the free-field can be decomposed into radial and angular components. Furthermore, the radial sound pressure in the far-field follows the \emph{inverse-distance law}~\cite{beranek2012acoustics}: $p\sim 1/r$, \hllg{as shown in Figure~\mbox{\ref{fig:inverse}}}. We utilize this property to extrapolate the full ASF from one of its far-field ``snapshots'' at a fixed radius, so that the full ASF does not need to be stored. Following the inverse-distance law, the sound pressure at any far-field location $(r,\theta,\phi)$ can be computed as
    \begin{equation}
    \label{eq:inversedist}
    p(r,\theta,\phi,\omega) = \frac{r_{ref}}{r}p(r_{ref},\theta,\phi,\omega),
    \end{equation}   
where $r_{ref}$ is the reference distance and only $p(r_{ref},\cdot,\cdot,\cdot)$ needs to be computed and stored. For brevity, we omit $r$ in following sections.

\subsubsection{Angular Pressure Field Encoding}
A spherical field consisting of a fixed number of points (e.g., $642$ points evenly distributed on a sphere surface) 
is obtained by generating an icosphere with 4 subdivisions. Real valued scattered sound pressures are evaluated at these field points during wave-based simulation. Spherical harmonics (SH) can represent a spherical scalar field compactly using a set of SH coefficients; they have been widely used for 3D sound field recording and reproduction~\cite{poletti2005three}. SH function up to order $l_{max}$ has $M=(l_{max}+1)^2$ coefficients. The angular pressure at the outgoing direction $(\theta, \phi)$ can be evaluated as
    \begin{equation}
    p(\theta, \phi, \omega)=\sum_{l=0}^{l_{max}} \sum_{m=-l}^{+l} Y_{l}^{m}(\theta, \phi)c_{l}^{m}(\omega),
    \label{eq:expansion}
    \end{equation} 
where $c_{l}^{m}(\omega)$ are the SH coefficients that encode our angular pressure fields. Increasing the number of coefficients can lead to more challenges because the dimension of our learning target is raised. 

\subsection{Learning Spherical Pressure Fields}
We need an appropriate geometric representation for the underlying objects in the scene so that we can apply geometric deep learning methods to compute the ASF. It is important that our approach should be able handle dynamic scenes with moving objects or changing topology. It can be difficult to handle such scenarios with mesh-based representations~\mbox{\cite{hanocka2019meshcnn,tan2018mesh, Zheng_2017_ICCV}}.
For example, \mbox{\cite{hanocka2019meshcnn}} calculates intrinsic geodesic distances for convolution operations, which cannot be applied when one big object breaks into two.

\begin{figure}[htbp]
  \includegraphics[width = \linewidth]{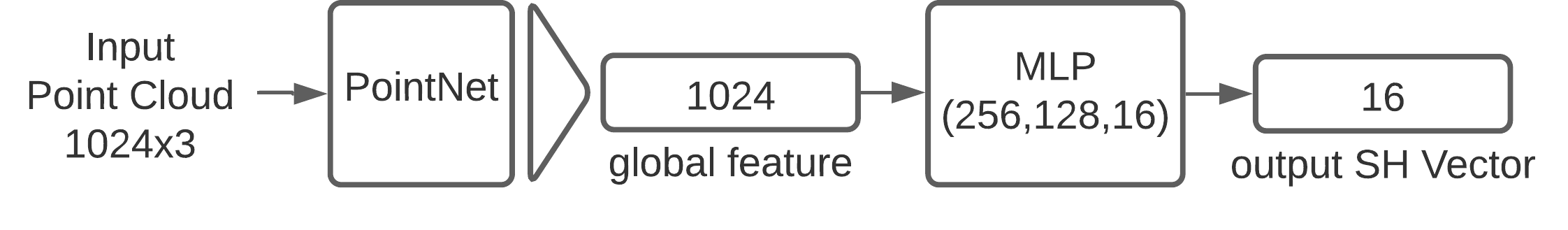}
  \caption{{\bf PointNet regression:} Given an input point cloud with $N=1024$ 3D points, we feed it to the PointNet architecture~\cite{pointnet} until maxpooling to extract the global feature. Then we use multi-layer perceptrons (MLPs) of layer size 256, 128, and 16 to map the feature to a SH vector of length 16 representing the scattering field.
  }
  \label{fig:pointnet}
  \vspace{-1em}
\end{figure}

Our approach uses a point cloud representation of the objects in the scene as an input. And we leverage the PointNet~\cite{pointnet} architecture to regress the spherical harmonics term $c_{l}^{m}$ in Equation~\ref{eq:expansion}. PointNet is a highly efficient and effective network architecture that works on raw point cloud input, and can perform various tasks including 3D object classification, semantic segmentation and our ASF regression. It also respects the permutation invariance of points. We slightly modify its output layers to predict the SH vector as shown in Figure~\ref{fig:pointnet}.

\section{Interactive Sound Propagation with Wave-Ray Coupling}

In this section, we describe how our learning-based method can be combined with geometric sound propagation techniques to compute the impulse responses for given  source and listener positions. Then, we can render them in highly dynamic scenes.

\paragraph{Hybrid Sound Propagation}
We use a hybrid sound propagation algorithm that combines wave-based and ray acoustics. Each of them handles different parts of wave acoustics phenomena, but they are coupled in terms of incoming and outgoing energies at multiple localized scattering fields. Specifically, our trained neural network estimates the scattering field and is used to compute propagation paths when sound interacts with obstacles in the scene. On the other hand, modeling sound propagation in the air along with specular and diffuse reflections at large boundary surfaces (e.g., walls, floors) is computed using ray tracing methods~\cite{schissler2014high,schissler2017interactive,rungta2018diffraction}.


\paragraph{Ray Tracing with Localized Fields}
Our localized ASFs are represented using SH coefficients. Given the most general ray tracing formulation at a scattering surface, the sound intensity $I_{out}$ of an outgoing direction $(\theta_o,\phi_o)$ from a scattering surface is given by the integral of the incoming intensity from all directions:
    \begin{equation}
    \label{eq:integral}
    I_{out}(\theta_o,\phi_o, \omega)=\int_{S} I_{in}(\theta_i, \phi_i, \omega)f(\theta_i, \phi_i, \theta_o, \phi_o, \omega) dS,
    \end{equation} 
where $S$ represents the directions on a spherical surface around the ray hit point, $I_{in}(\theta_i, \phi_i, \omega)$ is the incoming sound intensity from direction $(\theta_i,\phi_i)$, and $f(\theta_i, \phi_i, \theta_o, \phi_o, \omega)$ is the bi-directional scattering distribution function (BSDF) that is commonly used in visual rendering~\cite{pharr2016physically}. Our problem of acoustic wave scattering is different from visual rendering in two aspects: (1) sound wave scatters around objects, whereas light mostly transmits to visible directions or propagates through transparent materials; (2) BSDFs are point-based functions that depend on both incoming and outgoing directions, whereas our localized scattered fields are region-based functions. Therefore, we replace BSDFs in Equation~(\ref{eq:integral}) with our localized scattered field $p(\theta,\phi,\omega)$ representation from Equation~(\ref{eq:expansion}). Our choice of a spherical offset surface to model the scattered field also enables us to perform integration over the whole spherical surface in a straightforward manner, since evaluating spherical coordinates is efficient with SH functions. Although $p(\theta,\phi,\omega)$ encodes only the outgoing directions and assumes incoming plane waves to $-x$ direction, one can easily rotate the point cloud to align any incoming direction to the $-x$ direction and use our network to infer $p(\theta,\phi,\omega)$ at that direction. We update Equation~(\ref{eq:integral}) to
    \begin{equation}
    \label{eq:integral1}
    I_{out}(\theta_o,\phi_o, \omega)=\int_{S} I_{in}(\theta_i, \phi_i, \omega)p^2(\theta_i, \phi_i, \omega) dS.
    \end{equation} 
We  use the Monte Carlo integration to numerically evaluate the outgoing scattered intensity:
    \begin{equation}
    \label{eq:MC}
    I_{out}(\theta_o,\phi_o, \omega) \approx \frac{1}{N} \sum_{j=1}^{N} \frac{I_{in}(\theta_j,\phi_j, \omega)p^2(\theta_j, \phi_j, \omega)}{Pr(\theta_j,\phi_j)},
    \end{equation} 
where $N$ is the number of samples and $Pr(\theta_j,\phi_j)$ is the probability of generating a sample for direction $(\theta_j, \phi_j)$. A uniform sampling over the sphere surface gives $Pr(\theta_j,\phi_j)=\frac{1}{4\pi}$. 
\comment{In theory any probability distribution can be used.} 
As $N$ increases, the approximation becomes more accurate. 

\paragraph{Diffraction Compensation}
In wave acoustics, the total sound field at a position can be decomposed into the sum of the free-field sound pressure and the scattered sound field. Similar to~\cite{rungta2018diffraction}, we only have computed the scattered sound field up to now. But when the listener is obstructed from the sound source, the traditional ray-tracing algorithm will miss the contribution from the free-field, which will result in a very unnatural phenomenon: the sound would be greatly attenuated by a single obstacle if we only render the scattered sound, whereas in a realistic setup, low-frequency sound should not be attenuated by a small obstacle by much. To address this issue in a ray-tracing context, we propose to approximate sound interference with and without an obstacle depending on an extra visibility check. Specifically, for a sound source from direction $(\theta_j,\phi_j)$ and the listener at $(\theta_o,\phi_o)$, we calculate the sound at the listener position based on whether they are blocked by a scatterer from each other as:
\begin{equation}
\label{eq:hack}
I_{out}(\theta_o,\phi_o, \omega) \approx\left\{\begin{matrix}
\frac{1}{N} \sum_{j=1}^{N} \frac{I_{in}(\theta_j,\phi_j, \omega)(1-p^2(\theta_j, \phi_j, \omega))}{Pr(\theta_j,\phi_j)}, \text{if invisible}\\ 
 \frac{1}{N} \sum_{j=1}^{N} \frac{I_{in}(\theta_j,\phi_j, \omega)p^2(\theta_j, \phi_j, \omega)}{Pr(\theta_j,\phi_j)}, \text{if visible}
\end{matrix}\right.
\end{equation} 
Note that the visible case remains the same as Equation~\ref{eq:MC}, because the direct response will be automatically accounted for by the original ray-tracing pipeline. Obviously, this implementation is not physically accurate compared with wave acoustic simulations, since additional phase information is missing. However, this formulation will generate more realistic and more smooth sound rendering than prior work that only considers the scattering field, and we verify its benefits through a perceptual evaluation in $\S$~\ref{sec:study}.

\section{Implementation and Results}
In this section, we describe our implementation details and demonstrate the performance on many dynamic benchmarks.

\subsection{Data Generation}

\begin{figure}[htbp]
\centering
  \includegraphics[width=0.9\linewidth]{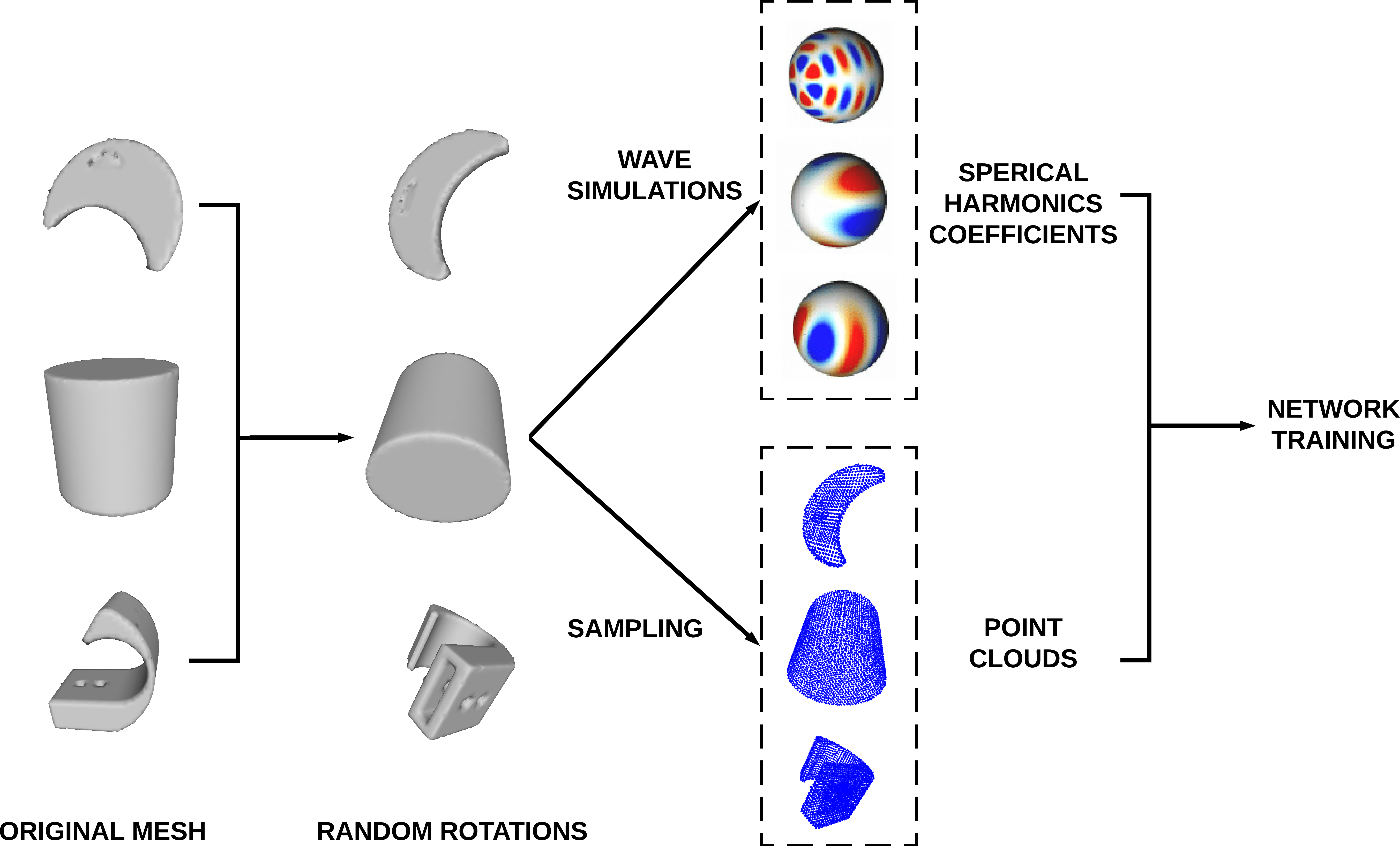}
  \caption{{\bf Our dataset generation pipeline for neural network training:} Given a set of CAD models, we apply random rotations with respect to their center of mass to generate a larger augmented dataset and  use a BEM solver to calculate the ASFs.}
  \label{fig:dataset}
  \vspace{-1em}
\end{figure}

\paragraph{Dataset}
To generate our learning examples, we choose to use the \emph{ABC Dataset}~\cite{Koch_2019_CVPR}. This dataset is a collection of one million general Computer-Aided Design (CAD) models  and is widely used for evaluation of geometric deep learning methods and applications. In particular, this dataset has been used to estimate of differential quantities (e.g., normals) and sharp features, which makes it attractive for learning ASFs as well. We sample \nObj models from the \emph{ABC Dataset} and process them by scaling objects such that their longest dimension \hllg{is in the range of $[1m, 2m]$}. The choice of such an object size limit is not fixed and can depend on the specific problem domain (e.g., size of objects used in applications like games or VR). Because the scattered pressure field is orientation-dependent, we augment our models by applying random 3D rotations to the original dataset to create an equal-sized rotation augmented dataset. To generate accurate labeled data, we use an accurate BEM wave solver, placing a plane wave source with unit strength propagating to the $-x$ direction. The solver outputs the ASF for each object, which becomes our learning target. The dataset pipeline is also illustrated in Figure~\ref{fig:dataset}.

\paragraph{Mesh Pre-processing}
The original meshes from the \emph{ABC Dataset} have high levels of details with fine edges of length shorter than $1cm$. Dense point cloud inputs could also be modeled or collected from the real-world scenes with granularity similar to this dataset. However, a high number of triangle elements in a mesh will significantly increase the simulation time of BEM solvers. For wave-based solver, our highest simulation frequency is \hllg{$1000Hz$}, which converts to a wavelength of \hllg{$34cm$}. Therefore, we use the standard procedure of mesh simplification and mesh clustering algorithm from the \emph{vcglib}~\footnote{\url{http://vcg.isti.cnr.it/vcglib/}} to ensure that our meshes have a minimum edge length of \hllg{$1.7cm$}, which is $1/20$ of our shortest target wavelength. This is sufficient according to the standard techniques used in BEM simulators~\cite{marburg2002six}. Most meshes after pre-processing have fewer than $20\%$ number of elements than the original and the BEM simulation for dataset generation gains over $10\times$ speedup. 

\paragraph{BEM Solver}
We use the \emph{FastBEM Acoustics} software~\footnote{\url{https://www.fastbem.com/}} as our wave-based solver. Simulations are run on a Windows 10 workstation that has 32 Intel(R) Xeon(R) Gold 5218 CPUs with multi-threading.  First we use the adaptive cross approximation (ACA) BEM~\cite{kurz2002adaptive} to compute the ASF since it can achieve near $\mathcal{O}(N)$ computational performance for small to medium sized models (e.g., element count $N\leq 100,000$). If it fails to converge within some fixed number of iterations, we use the conventional and accurate BEM solver. Overall, it takes about 12 days to compute the ASF up to $1000Hz$ frequency of about \nObj objects from the \emph{ABC Dataset}. The sound pressure field is evaluated at $642$ field points that are evenly distributed on the spherical field surface. Next, we use \emph{pyshtools}~\footnote{\url{https://shtools.oca.eu/shtools/public/index.html}} software~\cite{wieczorek2018shtools} to compute the spherical harmonics coefficients from  the pressure field using least squares inversion.

\paragraph{Reference Field Distance}
Since the inverse-distance law has increasing error in the near-field of objects, we need to find a suitable distance for computing our reference field. We experimentally simulate the sound pressure fall-off with respect to distance and observe that sound pressure that is $5m$ or further away from the scatterer closely agrees with this far-field approximation (see Figure~\ref{fig:inverse}). Therefore, we choose to calculate the pressure field on an offset surface $5m$ away from the scatterer's center using a BEM solver (i.e., setting $r_{ref}=5m$ in Equation~\ref{eq:inversedist}). \hllg{Note that this choice of $5m$ is not strict or fixed. If higher accuracy along the radial line is desired, multiple locations (especially in the near field) can be sampled during the simulation to interpolate the curve at a  higher accuracy. The precomputation time and memory overhead will increase linearly with respect to the number of sampled distance fields.}

\begin{figure}[htbp]
  \includegraphics[trim={0.5cm 0cm 3cm 2cm},clip,width=0.9\linewidth]{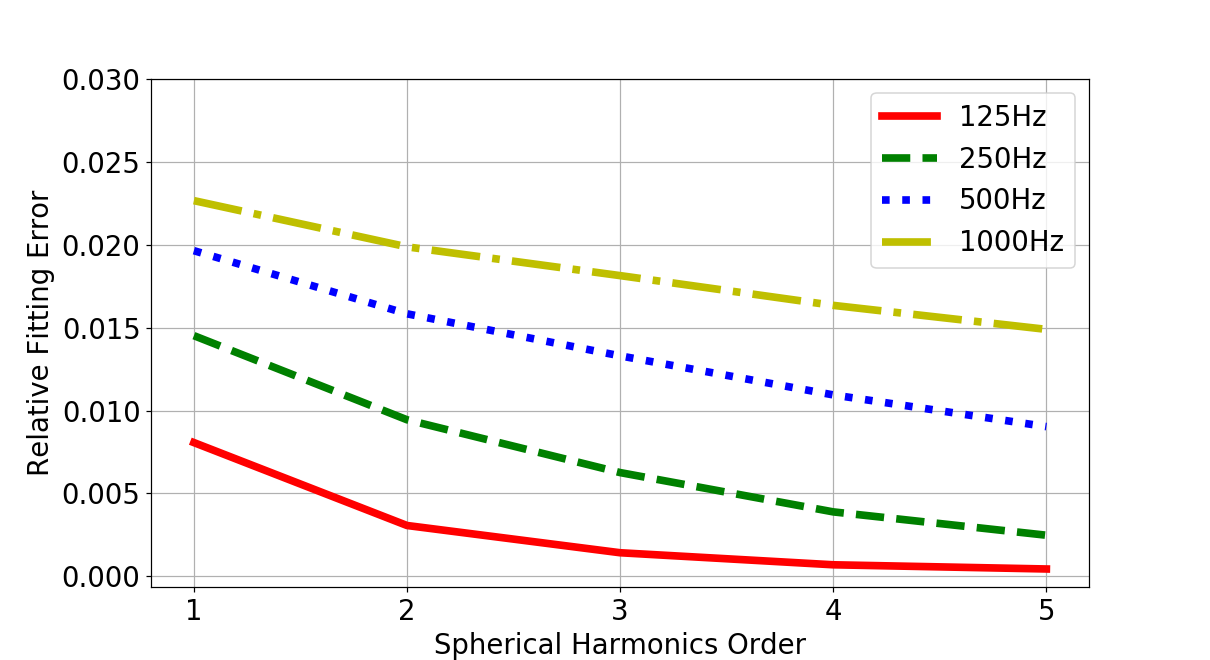}
  \caption{{\bf Spherical harmonics approximation of sound pressure fields:} We evaluate different orders of SH functions to fit our pressure fields at 4 frequencies and calculate the relative fitting errors.}
  \label{fig:sh}
  \vspace{-1em}
\end{figure}

\paragraph{Max Spherical Harmonics Order}
We experiment with the number of SH coefficients by projecting our scattered sound pressure fields to SH functions with different orders, as shown in Figure~\ref{fig:sh}. Based on this analysis, we choose to use up to a 3rd order SH projection, which yields sufficiently small fitting errors (relative error smaller than $2\%$) with $16$ SH coefficients. This sets the output of our neural network (Section 4.2.3) to be a vector of length $16$. 

\subsection{Network Training}
Our network model is trained on a GeForce RTX 2080 Ti GPU using the \emph{Tensorflow} framework~\cite{abadi2016tensorflow}. The dataset is split into training set and test set using the ratio $9:1$. \hllg{In the training stage, we use Adam optimizer to minimize $L_2$ norm loss between predicted spherical harmonic coefficients and the groundtruth. In practice, the initial learning rate is set to $1\times 10^{-3}$, which decays exponentially at a rate of 0.9 and clips at $1\times 10^{-5}$. The batch size is set to $128$ and typically our network converges after $100$ epochs in 8 hours. The number of our trainable parameters is about $800k$.}

\subsection{Runtime System and Benchmarks}
\label{sec:benchmarks}

We use the geometric sound propagation and rendering algorithm described in~\cite{schissler2014high}. \hllg{Our sound rendering system traces sound rays at octave frequency bands at $125Hz$, $250Hz$, $500Hz$, $1000Hz$, $2000Hz$, $4000Hz$, and $8000Hz$. The direct output from ray tracing for each frequency band is the energy histogram with respect to propagation delays. We take square root of these responses to compute the frequency dependent pressure response envelopes. Broadband frequency responses are interpolated from our traced frequency bands, and the inverse Fourier transform is used to re-construct the broadband impulse response. Our method does not preserve phase information, so a random phase spectrum is used during the inverse Fourier transform. In practice, this random spectrum does not introduce noticeable sound difference~\mbox{\cite{kuttruff1993auralization}}.}

\begin{table*}[!htbp]
\centering
\begin{tabularx}{\textwidth}{cXcc}
\toprule
Scene   &  Benchmark Description  & \#Triangle  & Frame time \\\hline
Floor   &  One static sound scatterer and one static sound source above an infinitely large floor. The listener moves horizontally so that the sound source visibility changes periodically. This is the simplest case where no sound reverberation occurs so as to accentuate the effect of sound diffraction.  &   4065     &     $10.65ms$       \\
Sibenik &  Two disjoint moving objects are used as scatterers in a church. The two scatterers revolve around each other in close proximity such that there are complicated near-field interactions of sound waves. This scene is a reverberant benchmark.  & 122798     &        $6.87ms$       \\
Trinity &  Six objects fly across a large indoor room and dynamically generate new composite scatterers or decompose into separate scatterers (i.e., changing topologies). As a result, the total number of separate scattering entities in the scene change and prior methods~\cite{rungta2018diffraction} are not effective. The occluded regions also change dynamically and create challenging scenarios for sound propagation.   & 386007    &   $12.95ms$ \\
Havana  &  Two rotating walls that are generally larger than scatterers in previous benchmarks in a half-open space. We use this benchmark to show that our approach can also handle large static objects, in addition to a large number of dynamic objects. It is an outdoor scene with moderate reverberation.  & 54383   &   $6.78ms$
\\\bottomrule
\end{tabularx}
\caption{Runtime performance on our benchmarks. The computation of ASFs takes $\leq 1ms$ per view and most frame time is spent in ray tracing.}
\label{tab:benchmarks}
\end{table*}

We require that the wall boundaries are explicitly marked in our scenes. As a result, when a ray hits the wall, only conventional sound reflections occur for all frequencies. During audio-visual rendering, when a ray hits a scattering object, we first extend the hit point along its ray direction by $0.5m$ and use it as the scattering region center. We include all the points within a search radius of $1m$ from the region center to generate a point cloud approximation of the scatterer. This point cloud is resampled using furthest point sampling and fed into our neural networks. Our network predicts the ASFs for sound frequencies corresponding to \hllg{$125Hz$, $250Hz$, $500Hz$ and $1000Hz$}. \hllg{The higher frequencies (i.e., $2000Hz$, $4000Hz$, and $8000Hz$) are handled by conventional geometric ray-tracing with specular and diffuse reflections and it does not use ASFs.} Our neural network has small prediction overhead of less than $1ms$ per view on an NVIDIA GeForce RTX 2080 Ti GPU. The interactive runtime propagation system is illustrated in Figure~\ref{fig:overview}. Our ray-tracer performs $200$ orders of reflections to generate late reverberation.

We evaluate the performance of our hybrid sound propagation and rendering algorithms several benchmark scenes shown in Figure~\ref{fig:teaser} and Table~\ref{sec:benchmarks}. They have with varying levels of dynamism in terms of moving objects and are demonstrated in our supplemental video. 

\subsection{Analysis}

\begin{figure*}[!ht]
    \subfloat[ASF of static objects from the unseen test set.]{\includegraphics[width=\textwidth]{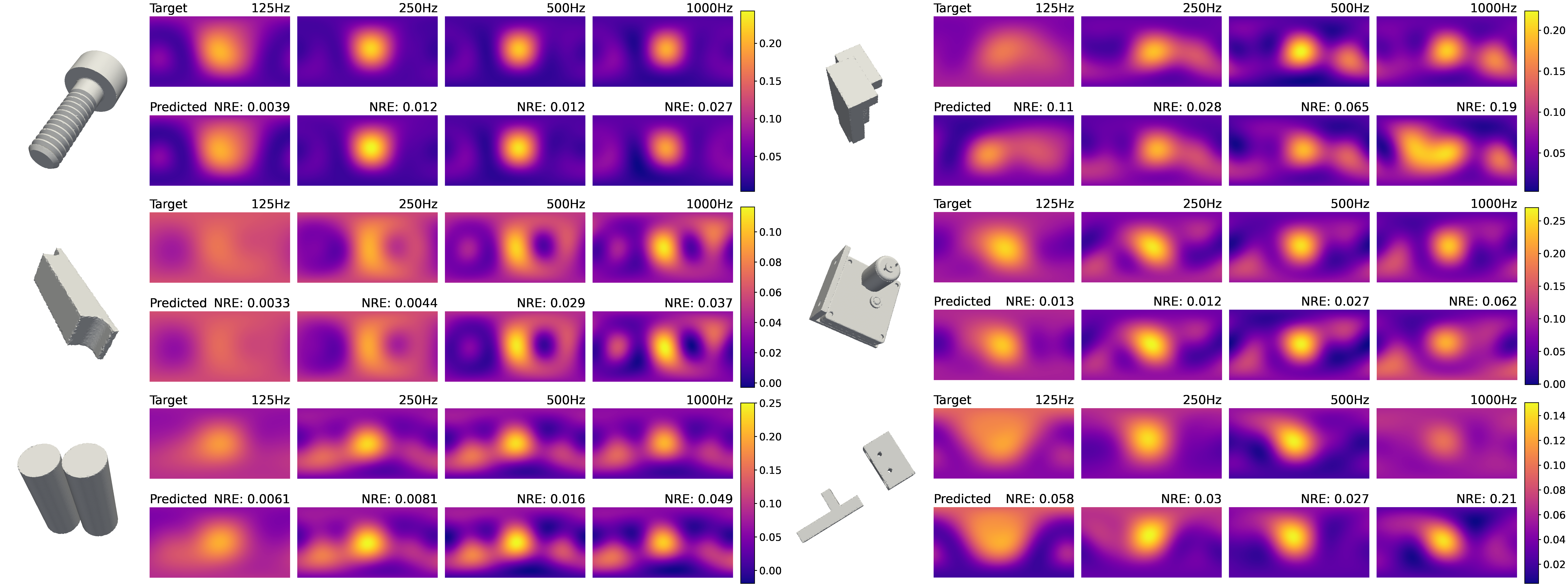} \label{fig:heatmap_static}}
    \vspace{-1em}
    \subfloat[ASF of dynamically moving objects (lowest and highest frequencies). We recompute the ASF at each time instance using our network. ]{\includegraphics[width=\textwidth]{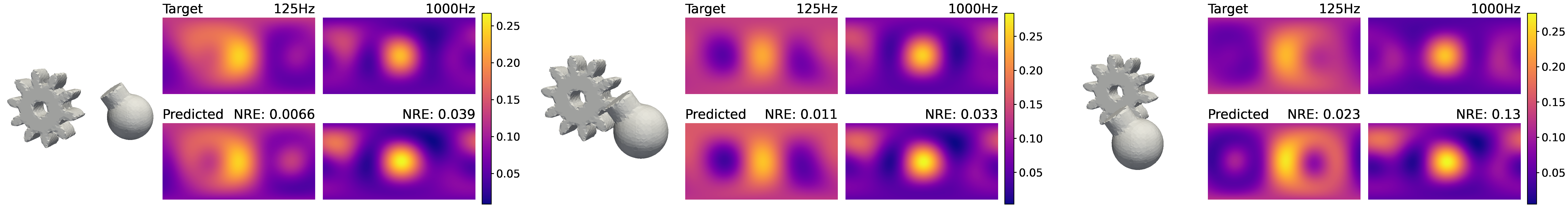} \label{fig:heatmap_dynamic}}
    \vspace{-1em}
    \subfloat[ASFs of a deforming object (lowest and highest frequencies), computed using our network.]{\includegraphics[width=\textwidth]{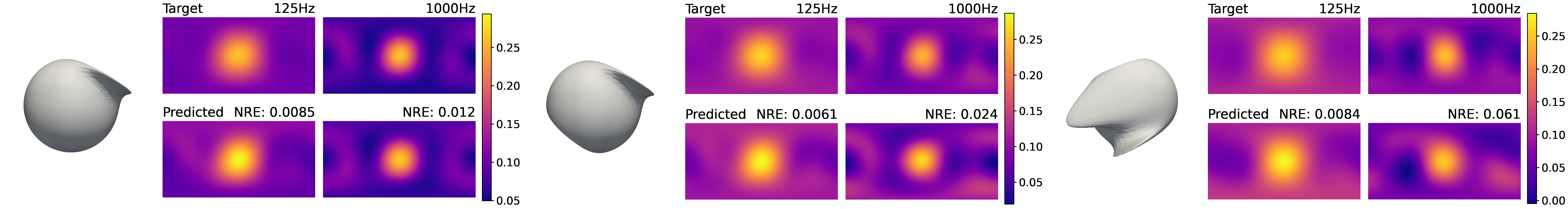} \label{fig:heatmap_deform}}
    \caption{{\bf Comparing ASF prediction accuracy in latitude-longitude plots:} We highlight the ASFs for different simulation frequencies. For each image block, the left column shows the mesh rendering of the objects. The Lat-Long plots visualize the ASF used in Equation (\ref{eq:integral}) by frequency using perceptually uniform colormaps: the top row (\emph{Target}) is the groundtruth ASF computed using a BEM solver on the original mesh; the bottom row (\emph{Predicted}) represents the ASF computed using our neural network based on point-cloud representation. The error metric NRE from Equation (\ref{eq:NRE}) is annotated above predicted ASFs. }
    \vspace{-1em}
    \label{fig:heatmap}
\end{figure*}

\paragraph{Accuracy Evaluation}
Our goal is to approximate the acoustic scattering fields of general 3D objects. While there is a preliminary 2D scattering dataset~\cite{fan2019fast}, there are no general or well-known datasets or benchmarks for evaluating such ASFs or related computations. Therefore, we use $10k$ objects from our test dataset to evaluate the performance of our trained network in terms of accuracy. 
Compared with the original \emph{ABC Dataset}, our test dataset has been augmented in terms of scale and using different orientations to evaluate the performance of our learning method. 
Since the prediction $p(\theta,\phi,\omega)\in [0,1]$ from our network is used as the BSDF in Equation (\ref{eq:integral}), by fixing $\omega$ and varying $\theta$ and $\phi$, we visualize the field using latitude-longitude plots in Figure~\ref{fig:heatmap}. 
We use the common normalized reproduction error (NRE)~\cite{lilis2010sound,betlehem2005theory} to measure the error level of our predicted fields, which is defined as:
\begin{equation}
    \label{eq:NRE}
    E(\omega)=\frac{\int_0^{2\pi}\int_0^{\pi}|p^{target}(\theta,\phi,\omega)-p^{predict}(\theta,\phi,\omega)|^2\mathrm{d}\phi\mathrm{d}\theta}{\int_0^{2\pi}\int_0^{\pi}|p^{target}(\theta,\phi,\omega)|^2\mathrm{d}\phi\mathrm{d}\theta}.
\end{equation} 

\begin{figure}[htbp]
\centering
  \includegraphics[width=\linewidth]{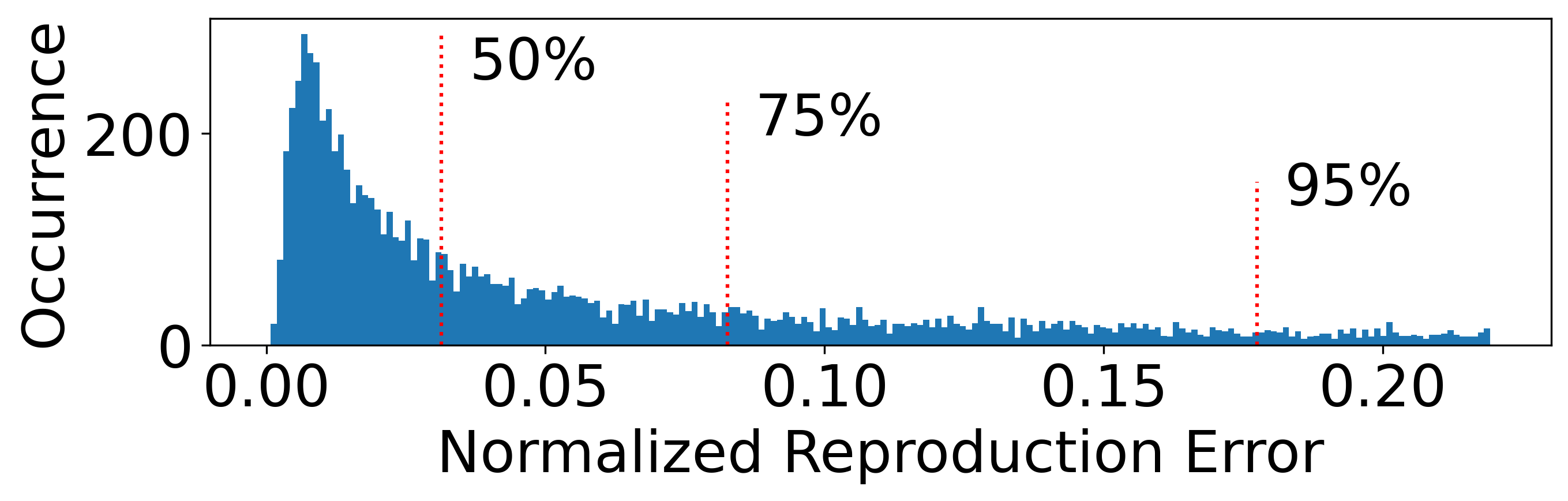}
  \caption{{\bf Distribution of test set prediction errors:} We also mark the $50\%,75\%$ and $95\%$ percentiles in the error histogram.}
  \label{fig:distribution}
  \vspace{-1em}
\end{figure}

\noindent We analyze three types of results. {\bf 1) Static Objects:} Figure \ref{fig:heatmap_static} shows a subset of CAD objects sampled from our test set, which is from the same distribution as the training set. The average NREs over the entire test set are $4.2\%, 7.6\%, 8.5\%, 10\%$ for $125Hz, 250Hz, 500Hz$, and $1000Hz$ respectively, with an overall NRE of $8.8\%$. In addition, we show the NRE distribution in Figure~\ref{fig:distribution}, where we see most test errors are contained below the average NRE. We observe a close visual match in most objects across frequencies. 
{\bf 2) Dynamic Objects:} Figure \ref{fig:heatmap_dynamic} shows an artificial example where two disjoint objects moves in proximity. Such scenarios are not created for the training set. We show the compraison and NREs at the lowest and highest frequencies. 
{\bf 3) Deforming objects:} Figure \ref{fig:heatmap_deform} shows an artificial example where one sphere undergoes deformation in different parts. 

These examples show that our network is able to perform consistently well on a large unseen test set when they are similar to the CAD models in training. Preliminary results on dynamic objects and deforming objects indicate that our network has the potential to generalize to more complicated scenarios that are not explicitly modeled during training, although we cannot provide the error bound on these cases. Note that the ASFs are not directly the perceived sound field at specific listener positions - instead they are intermediate transfer functions as one part in the sound rendering pipeline. Therefore, we further demonstrate the perceptual benefits of our predicted ASFs in Section~\ref{sec:study} and show that we can reliably generate plausible sound rendering under this error level. 


\paragraph{Frequency Growth}
In theory, our learning-based framework and runtime system can also incorporate wave frequencies beyond $1000Hz$. However, two important factors need to be considered when extending our setup: 1) the wave simulation time increases with the simulation frequency (e.g., between a square and cubic function for an accurate BEM solver); and 2) the ASF becomes more complicated at higher frequencies, which makes it more difficult to be learned or approximated using the same neural network. 
The per-object simulation time in our experiment is $0.87s, 1.10s, 2.04s, 2.80s$ for $125Hz, 250Hz, 500Hz$, and $1000Hz$, respectively. 
Note that the simulation time is governed much by the choice of the wave solver, as well as the relevant parameters/strategies used. We pre-processed our meshes according to the highest simulation frequency (i.e., the one with the shortest wavelength) and used that mesh representation for all frequencies. When a higher frequency needs to be added, the meshes need to have finer details, meaning more boundary elements will be involved (e.g., at least four times more elements when the simulation frequency doubles). A frequency-adaptive mesh simplification strategy~\cite{li2015interactive} can be used to reduce the simulation time at low frequencies. Our network prediction error also grows with the target frequency, but not at a prohibitive rate. We can reduce this error by using more training examples and more sophisticated neural network designs.

\section{Perceptual Evaluation}
\label{sec:study}

We perceptually evaluate our method using audio-visual listening tests. Our goal is to verify that our method generates plausible sound renderings and identify conditions it may or may not work well. We evaluate three sound rendering pipelines in our study: 1) Using predicted ASFs and our diffraction handling (ours); 2) Using predicted ASFs and the scattering sound rendering pipeline in diffraction kernels (DK)~\cite{rungta2018diffraction}; and 3) Using geometric sound propagation only (GSound)~\cite{schissler2017interactive}. 
The reason for choosing the two alternatives is that GSound is the state-of-the-art for interactive sound propagation without diffraction modeling. DK is regarded as state of the art hybrid algorithm  for interactive sound propagation in dynamic scenes with rigid objects and uses accurate ASFs precomputed using a BEM solver. Since wave-based methods are limited to static scenes, we do not include them in our perceptual evaluation.

\subsection{Participants}
We performed our studies using Amazon Mechanical Turk\footnote{https://www.mturk.com/} (AMT), a popular online crowdsourcing platform that can help data collection. We recruited 71 participants on AMT to take our study. To ensure the quality of our evaluation, we pre-screened our participants for this study. The pre-screening question is designed to test whether the participant has the proper listening device (e.g., a headphone) and is in a comfortable listening environment (i.e., not too noisy), so that they can tell basic qualitative differences between audios. Specifically, we convolved three impulse responses of reverberation times 0.2s, 0.6s and 1.0s with a 5-second long clean human speech recording to generate three corresponding reverberant speech. The commonly used just-noticeable-difference (JND) of reverberation times is a $5\%$ relative change~\cite{iso2009measurement}, so in normal conditions we expect a listener to correctly rank our three audios by their reverberation levels. Each participant is asked to listen to the three audios with no time limit, and sort them from the most reverberant/echoy to the least reverberant/echoy. The initial presentation order of the three audios is randomized for each participant. Out of the 71 participants who attempted our test, 51 participants qualified. After pre-screening, our participants consist of 35 males and 16 females, with an average age of 35.9 and a standard deviation of 9.5 years.

\subsection{Training}
\comment{Prior to the training, we ask participants to indicate their familiarity of the concept of sound diffraction. Among our participants, $22\%$ reported \emph{``I know nothing about it''}, $49\%$ reported \emph{``I have heard about it but am not sure''}, and $29\%$ reported \emph{``I learned about it and can describe what it is''}. As expected, general listeners have varied levels of understanding for sound effects, and we try to diminish this variance to some extent through a quick introduction of sound diffraction.}

During the training, we provide educational materials about sound diffraction including texts in non-academic language and a short YouTube video showing this phenomenon in the real world (where the sound travels around a pillar while the sound source is invisible). These materials require about one minute to read and watch but they are allowed to stay longer as needed. 

In addition, our participants become familiar with the video playing interface and are asked to adjust their audio playing volume to a comfortable level before the main listening tasks.

\subsection{Stimuli and Procedure}
We use the four scenes from benchmarks in $\S$\ref{sec:benchmarks} in combination with the three sound rendering pipelines to populate 12 audio-visual renderings that we ask our participants to give ratings on, with no time limit. We present the videos in four pages one after another, each page containing only three videos from the same scene (e.g., \emph{Floor (ours), Floor (DK)}, and \emph{Floor (GSound)}). The presentation order of the pages, as well as the order of videos within each page, have been randomized for each participant. Immediately after each video, participants are asked to give a sound reality rating and a sound smoothness rating. Both ratings are given from 0 to 5 stars, with a half-star granularity (i.e., there are a total 11 discrete levels). Participants are instructed to \emph{``give 5 stars for the most realistic and most smooth video and 0 star for the least realistic and smooth''}. Although we believe the standard of sound being realistic or smooth can vary among individuals, we expect that participants will be able to recognize cases where unnatural abrupt sound changes occur in response to scene dynamics, and will penalize them in their ratings.

\subsection{Results}
\begin{figure}[htbp]
\subfloat[Sound reality ratings by scene.]{%
  \includegraphics[clip,width=\columnwidth]{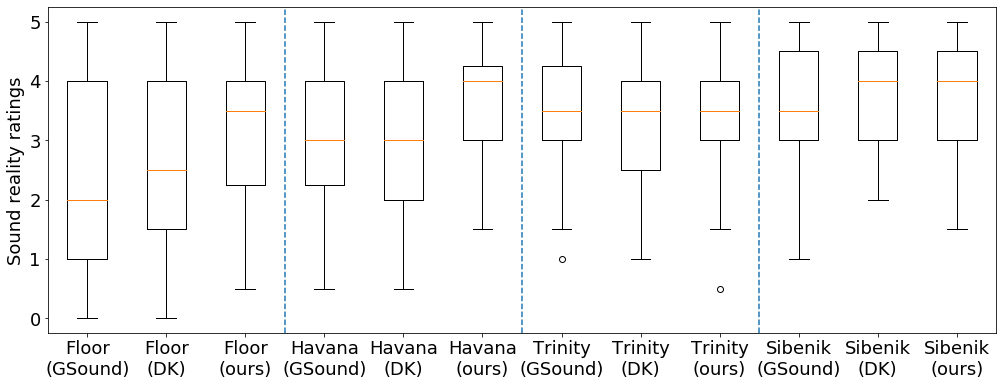}%
}

\subfloat[Sound smoothness ratings by scene.]{%
  \includegraphics[clip,width=\columnwidth]{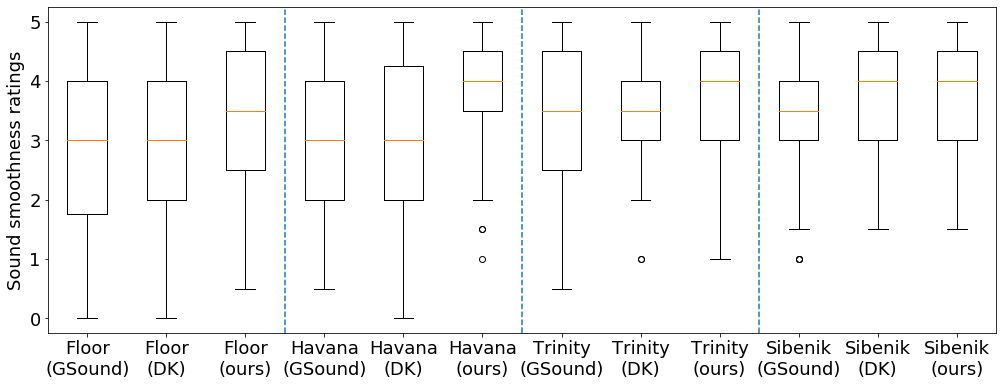}%
}

\caption{{\bf Perceptual evaluation results:} User ratings are visualized as box plots. A higher rating means better quality. Results are grouped by benchmark scene and each box represents the rating of a specific rendering pipeline in that scene.} 
\label{fig:user_study}
\end{figure}

The average study completion time is 13 minutes. We show the box plots of user ratings in Figure~\ref{fig:user_study}. We are interested in user's rating differences under the 3 test conditions (i.e., \emph{GSound, DK,} and \emph{ours}) on a per scene basis. Therefore, we perform within-group statistical analysis to identify potential significant differences. A significance level of 0.05 is adopted for all results in our discussions.

\paragraph{Sound Reality Ratings}
First we conduct a non-parametric Friedman test to the ratings given to the 3 rendering conditions, and find significant group differences in \emph{Floor} ($\chi^2 = 10.82, p < 0.01$) and \emph{Havana} ($\chi^2 = 8.27, p = 0.02$), but not in \emph{Trinity} ($\chi^2 = 0.16, p = 0.92$) or \emph{Sibenik} ($\chi^2 = 3.70, p = 0.16$). Note that \emph{Floor} and \emph{Havana} are basically open space scenes with less reverberation, whereas \emph{Trinity} and \emph{Sibenik} are common indoor environments that have a lot of reverberation. Considering that the sound power of reverberation is usually more dominant than diffraction, this result indicates that it is harder to tell the perceptual difference between these rendering pipelines when there is a strong reverberation. To identify the source of differences in \emph{Floor} and \emph{Havana} scenes, we perform post-hoc non-parametric Wilcoxon signed-rank tests with Bonferroni correction~\cite{holm1979simple}. We observe that \emph{ours} receives higher ratings than \emph{DK} and \emph{GSound} in both \emph{Floor} ($Z = \{215.0, 144.0\}, p < 0.01$) and \emph{Havana} ($Z=\{254.0, 186.5\}, p<0.01$). However, there are no significant differences between \emph{GSound} and \emph{DK} in any scene.

\paragraph{Sound Smoothness Ratings} 
Following the same procedure, we perform a Friedman test to the smoothness ratings, and discover that there are significant group differences in \emph{Floor} ($\chi^2=10.29,p<0.01$), \emph{Havana} ($\chi^2=7.63,p=0.02$), and \emph{Sibenik} ($\chi^2=12.59,p<0.01$). Post-hoc Wilcoxon tests show consistent results with reality ratings - we are only able to see a higher smoothness rating of \emph{ours} compared with both \emph{DK} and \emph{GSound} in \emph{Floor} ($Z=\{203.5,186.0\}, p=0.01$) and \emph{Havana} ($Z=\{233.5, 127.5\}, p<0.01$). In \emph{Sibenik}, both \emph{ours} and \emph{DK} receive a higher rating than \emph{GSound} ($Z=\{146.5, 171.0\}, p=0.01$). 

In conclusion, our pipeline receives better perceptual ratings than the other two methods in moderately reverberant conditions, which may not hold in highly reverberant scenes. We have increased perceptual differentiation over the \emph{DK} method. This is due to our better computation of the ASF for dynamic objects which \emph{DK} cannot handle well and our diffraction handling that aligns better with wave acoustic observations.

\section{Conclusions, Limitations, and Future Work}

We present a new learning-based approach to approximate the acoustic scattering fields of objects for interactive sound propagation. We exploit properties of the acoustic scattering field and use a geometric learning algorithm based on point-based approximation and the local shapes are encoded using implicit surfaces. We use a four-layer neural network that computes a field representation using 3rd order spherical harmonics. We use a large training dataset of \nObj objects, along with random 3D orientations and scaling of each object, and generate the accurate labeled data with a BEM solver. We evaluate the accuracy of our learning method on a large number of objects not seen in the training dataset, also undergoing topology changes. We observe low relative error in our benchmarks. Furthermore, we combine with a ray-tracing based sound propagation algorithm for sound rendering in highly dynamic scenes.
A perceptual study confirms that our approach generates smooth and realistic sound effects in  dynamic environments with increased perceptual differentiation over prior interactive methods.

Our approach has several limitations. These include all the challenges of geometric deep learning in terms of choosing an appropriate training dataset and long training time. Even though we observe an overall relative error of $8.8\%$ on thousands of new objects, it is very hard to provide any rigorous guarantees in terms of error bounds on arbitrary objects. Furthermore, we assume that objects in the scene are sound-hard and do not take into account various material properties. Our network has been tested for frequencies up to \hllg{$1000Hz$}, and we may need to design better learning methods for higher frequencies. The overall accuracy of our hybrid propagation algorithm lies between a pure geometric method and a global numeric solver. \hllg{There is a linear scaling of training time with the number of frequencies and the number of scattering objects, while the simulation time could scale as a cubic function of the frequency. As a result, the precomputation overhead can be high}. One mitigation is to limit the training to the kind of objects that are frequently used in an interactive application (e.g., a game or VR scenario). This equals to customized training for a specific application.

There are many avenues for future work. In addition to overcoming these limitations, we need to evaluate its performance in other scenarios and integrate with different applications. It would be useful to take into account the material properties by considering them as an additional object characteristic during training. We would also like to use other techniques from geometric processing and geometric deep learning to improve the performance of our approach. Our runtime ray tracing algorithm could use a different sampling scheme that exploits the properties of ASF.  In-person user study using a VR headset or standardized lab listening tests may add more insights to how spatial sound perception is affected by different sound propagation schemes.


\bibliographystyle{abbrv-doi}

\bibliography{sound}

\begin{thebibliography}{10}

\bibitem{SteamA}
Steam audio.
\newblock \url{https://valvesoftware.github.io/steam-audio}, 2018.

\bibitem{MicrosoftAcoustics}
Microsoft project acoustics.
\newblock \url{https://aka.ms/acoustics}, 2019.

\bibitem{OculusS}
Oculus spatializer.
\newblock
  \url{https://developer.oculus.com/downloads/package/oculus-spatializer-unity},
  2019.

\bibitem{abadi2016tensorflow}
M.~Abadi, P.~Barham, J.~Chen, Z.~Chen, A.~Davis, J.~Dean, M.~Devin,
  S.~Ghemawat, G.~Irving, M.~Isard, et~al.
\newblock Tensorflow: A system for large-scale machine learning.
\newblock In {\em 12th $\{$USENIX$\}$ Symposium on Operating Systems Design and
  Implementation ($\{$OSDI$\}$ 16)}, pp. 265--283, 2016.

\bibitem{beranek2012acoustics}
L.~L. Beranek and T.~Mellow.
\newblock {\em Acoustics: sound fields and transducers}.
\newblock Academic Press, 2012.

\bibitem{betlehem2005theory}
T.~Betlehem and T.~D. Abhayapala.
\newblock Theory and design of sound field reproduction in reverberant rooms.
\newblock {\em The Journal of the Acoustical Society of America},
  117(4):2100--2111, 2005.

\bibitem{botteldooren1995finite}
D.~Botteldooren.
\newblock Finite-difference time-domain simulation of low-frequency room
  acoustic problems.
\newblock {\em The Journal of the Acoustical Society of America},
  98(6):3302--3308, 1995.

\bibitem{chaitanya2019adaptive}
C.~R.~A. Chaitanya, J.~M. Snyder, K.~Godin, D.~Nowrouzezahrai, and
  N.~Raghuvanshi.
\newblock Adaptive sampling for sound propagation.
\newblock {\em IEEE transactions on visualization and computer graphics},
  25(5):1846--1854, 2019.

\bibitem{pointnet}
R.~Q. Charles, H.~Su, M.~Kaichun, and L.~J. Guibas.
\newblock Pointnet: Deep learning on point sets for 3d classification and
  segmentation.
\newblock {\em 2017 IEEE Conference on Computer Vision and Pattern Recognition
  (CVPR)}, Jul 2017. doi: {{%
10\hspace{.1pt}\discretionary{.}{%
}{.}\hspace{.4pt}1109\discretionary{/}{%
}{/}cvpr\hspace{.1pt}\discretionary{.}{%
}{.}\hspace{.4pt}2017\hspace{.1pt}\discretionary{.}{%
}{.}\hspace{.4pt}16}}


\bibitem{eaton2016estimation}
J.~Eaton, N.~D. Gaubitch, A.~H. Moore, P.~A. Naylor, J.~Eaton, N.~D. Gaubitch,
  A.~H. Moore, P.~A. Naylor, N.~D. Gaubitch, J.~Eaton, et~al.
\newblock Estimation of room acoustic parameters: The ace challenge.
\newblock {\em IEEE/ACM Transactions on Audio, Speech and Language Processing
  (TASLP)}, 24(10):1681--1693, 2016.

\bibitem{fan2019fast}
Z.~Fan, V.~Vineet, H.~Gamper, and N.~Raghuvanshi.
\newblock Fast acoustic scattering using convolutional neural networks.
\newblock In {\em ICASSP 2020-2020 IEEE International Conference on Acoustics,
  Speech and Signal Processing (ICASSP)}, pp. 171--175. IEEE, 2020.

\bibitem{ferguson2018sound}
E.~L. Ferguson, S.~B. Williams, and C.~T. Jin.
\newblock Sound source localization in a multipath environment using
  convolutional neural networks.
\newblock In {\em 2018 IEEE International Conference on Acoustics, Speech and
  Signal Processing (ICASSP)}, pp. 2386--2390. IEEE, 2018.

\bibitem{funkhouser1998}
T.~Funkhouser, I.~Carlbom, G.~Elko, G.~Pingali, M.~Sondhi, and J.~West.
\newblock A beam tracing approach to acoustic modeling for interactive virtual
  environments.
\newblock In {\em Proceedings of the 25th annual conference on Computer
  graphics and interactive techniques}, pp. 21--32. ACM, 1998.

\bibitem{genovese2019blind}
A.~F. Genovese, H.~Gamper, V.~Pulkki, N.~Raghuvanshi, and I.~J. Tashev.
\newblock Blind room volume estimation from single-channel noisy speech.
\newblock In {\em ICASSP 2019-2019 IEEE International Conference on Acoustics,
  Speech and Signal Processing (ICASSP)}, pp. 231--235. IEEE, 2019.

\bibitem{hanocka2019meshcnn}
R.~Hanocka, A.~Hertz, N.~Fish, R.~Giryes, S.~Fleishman, and D.~Cohen-Or.
\newblock Meshcnn: a network with an edge.
\newblock {\em ACM Transactions on Graphics (TOG)}, 38(4):1--12, 2019.

\bibitem{holm1979simple}
S.~Holm.
\newblock A simple sequentially rejective multiple test procedure.
\newblock {\em Scandinavian journal of statistics}, pp. 65--70, 1979.

\bibitem{iso2009measurement}
A.~ISO.
\newblock Measurement of room acoustic parameters - part 1.
\newblock {\em ISO Std}, 2009.

\bibitem{james2006precomputed}
D.~L. James, J.~Barbi{\v{c}}, and D.~K. Pai.
\newblock Precomputed acoustic transfer: output-sensitive, accurate sound
  generation for geometrically complex vibration sources.
\newblock In {\em ACM Transactions on Graphics (TOG)}, vol.~25, pp. 987--995.
  ACM, 2006.

\bibitem{Koch_2019_CVPR}
S.~Koch, A.~Matveev, Z.~Jiang, F.~Williams, A.~Artemov, E.~Burnaev, M.~Alexa,
  D.~Zorin, and D.~Panozzo.
\newblock Abc: A big cad model dataset for geometric deep learning.
\newblock In {\em The IEEE Conference on Computer Vision and Pattern
  Recognition (CVPR)}, June 2019.

\bibitem{krokstad1968}
A.~Krokstad, S.~Strom, and S.~S{\o}rsdal.
\newblock Calculating the acoustical room response by the use of a ray tracing
  technique.
\newblock {\em Journal of Sound and Vibration}, 8(1):118--125, 1968.

\bibitem{kurz2002adaptive}
S.~Kurz, O.~Rain, and S.~Rjasanow.
\newblock The adaptive cross-approximation technique for the 3d
  boundary-element method.
\newblock {\em IEEE transactions on Magnetics}, 38(2):421--424, 2002.

\bibitem{kuttruff1993auralization}
K.~H. Kuttruff.
\newblock Auralization of impulse responses modeled on the basis of ray-tracing
  results.
\newblock {\em Journal of the Audio Engineering Society}, 41(11):876--880,
  1993.

\bibitem{larsson2002}
P.~Larsson, D.~Vastfjall, and M.~Kleiner.
\newblock Better presence and performance in virtual environments by improved
  binaural sound rendering.
\newblock In {\em Virtual, Synthetic, and Entertainment Audio conference}, Jun
  2002.

\bibitem{lauterbach2007}
C.~Lauterbach, A.~Chandak, and D.~Manocha.
\newblock Interactive sound rendering in complex and dynamic scenes using
  frustum tracing.
\newblock {\em IEEE Transactions on Visualization and Computer Graphics},
  13(6):1672--1679, 2007.

\bibitem{li2015interactive}
D.~Li, Y.~Fei, and C.~Zheng.
\newblock Interactive acoustic transfer approximation for modal sound.
\newblock {\em ACM Transactions on Graphics (TOG)}, 35(1):1--16, 2015.

\bibitem{lilis2010sound}
G.~N. Lilis, D.~Angelosante, and G.~B. Giannakis.
\newblock Sound field reproduction using the lasso.
\newblock {\em IEEE Transactions on Audio, Speech, and Language Processing},
  18(8):1902--1912, 2010.

\bibitem{marburg2002six}
S.~Marburg.
\newblock Six boundary elements per wavelength: Is that enough?
\newblock {\em Journal of computational acoustics}, 10(01):25--51, 2002.

\bibitem{mehra2013wave}
R.~Mehra, N.~Raghuvanshi, L.~Antani, A.~Chandak, S.~Curtis, and D.~Manocha.
\newblock Wave-based sound propagation in large open scenes using an equivalent
  source formulation.
\newblock {\em ACM Transactions on Graphics (TOG)}, 32(2):19, 2013.

\bibitem{mehra2015wave}
R.~Mehra, A.~Rungta, A.~Golas, M.~Lin, and D.~Manocha.
\newblock Wave: Interactive wave-based sound propagation for virtual
  environments.
\newblock {\em IEEE transactions on visualization and computer graphics},
  21(4):434--442, 2015.

\bibitem{pharr2016physically}
M.~Pharr, W.~Jakob, and G.~Humphreys.
\newblock {\em Physically based rendering: From theory to implementation}.
\newblock Morgan Kaufmann, 2016.

\bibitem{pierce1990acoustics}
A.~D. Pierce and R.~T. Beyer.
\newblock Acoustics: An introduction to its physical principles and
  applications. 1989 edition, 1990.

\bibitem{poletti2005three}
M.~A. Poletti.
\newblock Three-dimensional surround sound systems based on spherical
  harmonics.
\newblock {\em Journal of the Audio Engineering Society}, 53(11):1004--1025,
  2005.

\bibitem{pulkki2019machine}
V.~Pulkki and U.~P. Svensson.
\newblock Machine-learning-based estimation and rendering of scattering in
  virtual reality.
\newblock {\em The Journal of the Acoustical Society of America},
  145(4):2664--2676, 2019.

\bibitem{raghuvanshi2009efficient}
N.~Raghuvanshi, R.~Narain, and M.~C. Lin.
\newblock Efficient and accurate sound propagation using adaptive rectangular
  decomposition.
\newblock {\em IEEE Transactions on Visualization and Computer Graphics},
  15(5):789--801, 2009.

\bibitem{nikunj2014}
N.~Raghuvanshi and J.~Snyder.
\newblock Parametric wave field coding for precomputed sound propagation.
\newblock {\em ACM Transactions on Graphics (TOG)}, 33(4):38, 2014.

\bibitem{raghuvanshi2018parametric}
N.~Raghuvanshi and J.~Snyder.
\newblock Parametric directional coding for precomputed sound propagation.
\newblock {\em ACM Transactions on Graphics (TOG)}, 37(4):108, 2018.

\bibitem{Raghuvanshi:2010:PWS}
N.~Raghuvanshi, J.~Snyder, R.~Mehra, M.~Lin, and N.~Govindaraju.
\newblock Precomputed wave simulation for real-time sound propagation of
  dynamic sources in complex scenes.
\newblock {\em ACM Trans. Graph.}, 29(4):68:1--68:11, July 2010.

\bibitem{rosen}
M.~Rosen, K.~W. Godin, and N.~Raghuvanshi.
\newblock {Interactive Sound Propagation For Dynamic Scenes Using 2d Wave
  Simulation}.
\newblock {\em Computer Graphics Forum}, 2020. doi: {{%
10\hspace{.1pt}\discretionary{.}{%
}{.}\hspace{.4pt}1111\discretionary{/}{%
}{/}cgf\hspace{.1pt}\discretionary{.}{%
}{.}\hspace{.4pt}14099}}


\bibitem{rungta2016psychoacoustic}
A.~Rungta, S.~Rust, N.~Morales, R.~Klatzky, M.~Lin, and D.~Manocha.
\newblock Psychoacoustic characterization of propagation effects in virtual
  environments.
\newblock {\em ACM Transactions on Applied Perception (TAP)}, 13(4):21, 2016.

\bibitem{rungta2018diffraction}
A.~Rungta, C.~Schissler, N.~Rewkowski, R.~Mehra, and D.~Manocha.
\newblock Diffraction kernels for interactive sound propagation in dynamic
  environments.
\newblock {\em IEEE Transactions on Visualization and Computer Graphics},
  24(4):1613--1622, 2018.

\bibitem{savioja2015overview}
L.~Savioja and U.~P. Svensson.
\newblock Overview of geometrical room acoustic modeling techniques.
\newblock {\em The Journal of the Acoustical Society of America},
  138(2):708--730, 2015.

\bibitem{schissler2017interactive}
C.~Schissler and D.~Manocha.
\newblock Interactive sound propagation and rendering for large multi-source
  scenes.
\newblock {\em ACM Transactions on Graphics (TOG)}, 36(1):2, 2017.

\bibitem{schissler2018interactive}
C.~Schissler and D.~Manocha.
\newblock Interactive sound rendering on mobile devices using ray-parameterized
  reverberation filters.
\newblock {\em arXiv preprint arXiv:1803.00430}, 2018.

\bibitem{schissler2014high}
C.~Schissler, R.~Mehra, and D.~Manocha.
\newblock High-order diffraction and diffuse reflections for interactive sound
  propagation in large environments.
\newblock {\em ACM Transactions on Graphics (TOG)}, 33(4):39, 2014.

\bibitem{svensson1999}
U.~P. Svensson, R.~I. Fred, and J.~Vanderkooy.
\newblock An analytic secondary source model of edge diffraction impulse
  responses.
\newblock {\em The Journal of the Acoustical Society of America},
  106(5):2331--2344, 1999.

\bibitem{tan2018mesh}
Q.~Tan, L.~Gao, Y.-K. Lai, J.~Yang, and S.~Xia.
\newblock Mesh-based autoencoders for localized deformation component analysis.
\newblock In {\em Thirty-Second AAAI Conference on Artificial Intelligence},
  2018.

\bibitem{tang2019scene}
Z.~Tang, N.~J. Bryan, D.~Li, T.~R. Langlois, and D.~Manocha.
\newblock Scene-aware audio rendering via deep acoustic analysis.
\newblock {\em IEEE Transactions on Visualization and Computer Graphics}, 2020.

\bibitem{Taylor12}
M.~Taylor, A.~Chandak, Q.~Mo, C.~Lauterbach, C.~Schissler, and D.~Manocha.
\newblock Guided multiview ray tracing for fast auralization.
\newblock {\em IEEE Transactions on Visualization and Computer Graphics},
  18:1797--1810, 2012.

\bibitem{tenenbaum2019room}
R.~A. Tenenbaum, F.~O. Taminaro, and V.~Melo.
\newblock Room acoustics modeling using a hybrid method with fast auralization
  with artificial neural network techniques.
\newblock In {\em Proc. International Congress on Acoustics (ICA)}, pp.
  6420--6427, 2019.

\bibitem{thompson2006review}
L.~L. Thompson.
\newblock A review of finite-element methods for time-harmonic acoustics.
\newblock {\em The Journal of the Acoustical Society of America},
  119(3):1315--1330, 2006.

\bibitem{tsingos2009precomputing}
N.~Tsingos.
\newblock Precomputing geometry-based reverberation effects for games.
\newblock In {\em Audio Engineering Society Conference: 35th International
  Conference: Audio for Games}. Audio Engineering Society, 2009.

\bibitem{tsingos2001}
N.~Tsingos, T.~Funkhouser, A.~Ngan, and I.~Carlbom.
\newblock Modeling acoustics in virtual environments using the uniform theory
  of diffraction.
\newblock In {\em Proceedings of the 28th annual conference on Computer
  graphics and interactive techniques}, pp. 545--552. ACM, 2001.

\bibitem{tsokaktsidis2019artificial}
D.~Tsokaktsidis, T.~Von~Wysocki, F.~Gauterin, and S.~Marburg.
\newblock Artificial neural network predicts noise transfer as a function of
  excitation and geometry.
\newblock In {\em Proc. International Congress on Acoustics (ICA)}, pp.
  4392--4396, 2019.

\bibitem{valimaki2012fifty}
V.~Valimaki, J.~D. Parker, L.~Savioja, J.~O. Smith, and J.~S. Abel.
\newblock Fifty years of artificial reverberation.
\newblock {\em IEEE Transactions on Audio, Speech, and Language Processing},
  20(5):1421--1448, 2012.

\bibitem{vorlander1989}
M.~Vorl{\"a}nder.
\newblock Simulation of the transient and steady-state sound propagation in
  rooms using a new combined ray-tracing/image-source algorithm.
\newblock {\em The Journal of the Acoustical Society of America},
  86(1):172--178, 1989.

\bibitem{wieczorek2018shtools}
M.~A. Wieczorek and M.~Meschede.
\newblock Shtools: Tools for working with spherical harmonics.
\newblock {\em Geochemistry, Geophysics, Geosystems}, 19(8):2574--2592, 2018.

\bibitem{wrobel2003boundary}
L.~C. Wrobel and A.~Kassab.
\newblock Boundary element method, volume 1: Applications in thermo-fluids and
  acoustics.
\newblock {\em Appl. Mech. Rev.}, 56(2):B17--B17, 2003.

\bibitem{yeh2013wave}
H.~Yeh, R.~Mehra, Z.~Ren, L.~Antani, D.~Manocha, and M.~Lin.
\newblock Wave-ray coupling for interactive sound propagation in large complex
  scenes.
\newblock {\em ACM Transactions on Graphics (TOG)}, 32(6):165, 2013.

\bibitem{Zheng_2017_ICCV}
X.~Zheng, C.~Wen, N.~Lei, M.~Ma, and X.~Gu.
\newblock Surface registration via foliation.
\newblock In {\em The IEEE International Conference on Computer Vision (ICCV)},
  Oct 2017.

\end{thebibliography}

\pagebreak
\clearpage

\appendix
\appendixpage
\section{Wave Acoustics and the Helmholtz Equation}
\label{sec:append_helmholtz}
A scalar acoustic pressure field, $P(\boldx, t)$, satisfies the homogeneous wave equation
\begin{equation}
\label{eq:waveequation}
\nabla ^2 P - \frac{1}{c^2} \frac{ \partial^2 P}{\partial t ^2 } = 0,
\end{equation}
where $c$ is the speed of sound. We can analyze the pressure field in the frequency domain using Fourier transform
\begin{equation}
\label{eq:fourier}
p(\boldx, \omega)=\mathcal{F}_{t}\{P(\boldx, t)\}=\int_{-\infty}^{\infty} P(\boldx, t) e^{-j \omega t} dt.
\end{equation}
At each frequency $\omega$ the pressure field satisfies the homogeneous Helmholtz wave equation
\begin{equation}
\label{eq:helmholtz}
(\nabla^{2} + k^{2}) p(\boldx, \omega)=0,
\end{equation}
where $k=\frac{\omega}{c}$ is the wavenumber. We can expand the Laplacian operator in terms of spherical coordinates $(r,\theta,\phi)$ as
\begin{equation}
\label{eq:spherical}
\left(\frac{\partial^{2}}{\partial r^{2}}+\frac{2}{r} \frac{\partial}{\partial r}+\frac{1}{r^{2} \sin \theta} \frac{\partial}{\partial \theta}\left(\sin \theta \frac{\partial}{\partial \theta}\right)+\frac{1}{r^{2} \sin ^{2} \theta} \frac{\partial^{2}}{\partial \phi^{2}}+k^2\right) p=0.
\end{equation}
The general free-field solution of (\ref{eq:spherical}) can be formulated as
\begin{equation}
\label{eq:solution}
p(\boldx, \omega)=\sum_{l=0}^{\infty} \sum_{m=-l}^{+l}\left[A_{l m} h_{l}^{(1)}(k r)+B_{l m} h_{l}^{(2)}(k r)\right] Y_{l}^{m}(\theta, \phi),
\end{equation}
where $h_{l}^{(1)}$ and $h_{l}^{(2)}$ are Hankel functions of the first and the second kind, respectively. $A_{l m}$ and $B_{l m}$ are arbitrary constants, $A_{l m} h_{l}^{(1)}(k r)+B_{l m} h_{l}^{(2)}(k r)$ together represents the radial part of the solution and the spherical harmonics term $Y_{l}^{m}(\theta, \phi)$ represents the angular part of the solution. 

\section{Acoustic Wave Scattering}
\label{sec:append_scattering}
Equation~(\ref{eq:helmholtz}) describes the behavior of acoustic waves in free-field conditions. When a propagating acoustic wave generated by a sound source interacts with an obstacle (the scatterer), a scattered field is generated outside the scatterer. The  Helmholtz equation can be used to describe this scenario:
    \begin{equation}
    \label{eq:inhomogeneous}
    (\nabla^{2} + k^{2}) p(\boldx, \omega)=-Q(\boldx,\omega),\quad \forall\boldx\in E,
    \end{equation}
where $E$ is the space that is exterior to the scatterer and $Q(\boldx,\omega)$ represents the acoustic sources in the frequency domain. Common types of sound sources include monopole sources, dipole sources, and plane wave sources. To obtain an exact solution to~(\ref{eq:inhomogeneous}), the boundary conditions on the scatterer surface $S$ need to be specified. In this work, we assume all the scattering objects are sound-hard (i.e. all energy is scattered, not absorbed) and therefore use the zero Neumann boundary condition for all $S$:
    \begin{equation}
    \label{eq:neumann}
    \frac{\partial p}{\partial \mathbf{n}(\boldx)}=0,\quad \forall\boldx\in S,
    \end{equation}    
where $\mathbf{n}(\boldx)$ is the normal vector at $\boldx$. Alternatively, other conditions including the sound-soft Dirichlet boundary condition and the mixed Robin boundary condition~\cite{pierce1990acoustics} can be used to model different acoustic scattering problems. When the boundary conditions are fully defined, the constants in Equation~\ref{eq:solution} can be uniquely determined.

\section{Comparison with prior Hybrid Schemes}
Our hybrid pipeline is similar to the diffraction-kernel method (DK)~\cite{rungta2018diffraction}. We compute the acoustic scattering fields (ASFs) for each object in the scene, which tend to capture all the interactions betweeen the sound waves with the object. These include reflections, diffraction, scattering and interference. In particular, the ASFs map the incoming sound field reaching the object to outgoing, diffracted field that emanates from the object. At runtime, these ASFs are integrated with  ray tracing to compute the specular and diffuse reflections at interactive rates.

The DK algorithm essentially precomputes the ASF for each object using a BEM solver and relies on the symmetry of an object to save precomputation time. On the other hand, our learning-based algorithm approximates the ASFs using a neural network at runtime and performs GPU computations. Our approach is designed to overcome two main limitations of DK:
\begin{enumerate}
    \item DK is limited to rigid objects and precomputes the exact acoustic scattering field (ASF) using an accurate wave-solver like BEM. However, if the object undergoes any small deformation at runtime, the symmetry of the object breaks and DK is no longer effective (full simulation required). Instead, our learning method can compute a new approximation of the ASF using our neural network for deforming objects.
    
    \item  DK assumes that the known rigid objects are well-separated at runtime. If two objects are in close proximity or ``glue'' together (as seen in Sibenik and Trinity benchmarks in Table 1), DK will not work. Instead, our learning method can approximate the ASF using our neural network.  
\end{enumerate}
At the same time, our learning-based algorithm only approximates the ASF using a neural-network. The accuracy of our learning method can change depending on the training dataset, the loss functions and hyperparameters used by the network. We have tested the performance on thousands of unseen objects in the ABC Dataset and observed $8.8\%$ overall error in the resulting pressure field (see Fig.\ref{fig:heatmap}). However, it is hard to provide any rigorous guarantees on the maximum error in the ASF for an arbitrary object. This is a fundamental limitation of machine learning methods.

In terms of precomputation costs and runtime costs, our learning-based method is slightly more expensive than DK.

\noindent {\bf Precomputation Cost:}  Our precomputation cost is governed by data generation process, as explained in Section 6.1. We compute the exact ASF of these objects using BEM solver (which is similar to computing the diffraction kernel). However, we perform the additional step of network training, as explained in Section 6.2, which can take 8 hours for each frequency. After this training step, we only store the network, and do not need to store the ASFs. On the other hand, DK may only compute the ASFs of all the rigid objects that are used in an application, when there is no demand for adding new objects.

\noindent {\bf Runtime Cost:} Most of the runtime cost is dominated by ray tracing, which is almost the same for [DK] and our method. While DK uses precomputed ASFs of rigid objects, we approximate the new ASF of an object using the network, which takes less than 1ms on NVIDIA GPU (as explained Section 6.3). This runtime GPU computation is an ignorable overhead of our learning method, although it poses an extra hard-ware requirement.

\end{document}